# Twist-angle and thickness-ratio tuning of plasmon polaritons in twisted bilayer van der Waals films

Chong Wang[1,2+], Yuangang Xie[3+], Junwei Ma[3], Guangwei Hu[5], Qiaoxia Xing[3], Shenyang Huang[3], Chaoyu Song[3], Fanjie Wang[3], Yuchen Lei[3], Jiasheng Zhang[3], Lei Mu[3], Tan Zhang[4], Yuan Huang[6], Cheng-Wei Qiu[4]*, Yugui Yao[1,2]*, Hugen Yan[3]*

1, Centre for Quantum Physics, Key Laboratory of Advanced Optoelectronic Quantum Architecture and Measurement (MOE), School of Physics, Beijing Institute of Technology, Beijing, 100081, China.
2, Beijing Key Lab of Nanophotonics & Ultrafine Optoelectronic Systems, School of Physics, Beijing Institute of Technology, Beijing, 100081, China.
3, State Key Laboratory of Surface Physics, Key Laboratory of Micro and Nano-Photonic Structures (Ministry of Education), and Department of Physics, Fudan University, Shanghai 200433, China.
4, Department of Electrical and Computer Engineering, National University of Singapore, Singapore 117583, Singapore
5, School of Electrical and Electronic Engineering, 50 Nanyang Avenue, Nanyang Technological University, Singapore, 639798, Singapore.
6, Advanced Research Institute of Multidisciplinary Science, Beijing Institute of Technology, Beijing, 100081, China

+These authors contributed equally to this work.

*Corresponding author. E-mail: hgyan@fudan.edu.cn (H. Y.), ygyao@bit.edu.cn (Y. Y.), eleqc@nus.edu.sg (Q. C.)

## Abstract:

Stacking bilayer structures is an efficient way to tune the topology of polaritons in in-plane anisotropic films, e.g., by leveraging the twist angle (TA). However, the effect of another geometric parameter, film thickness ratio (TR), on manipulating the plasmon topology in bilayers is elusive. Here, we fabricate bilayer structures of $WTe_2$ films, which naturally host in-plane hyperbolic plasmons in the terahertz range. Plasmon topology is successfully modified by changing the TR and TA synergistically, manifested by the extinction spectra of unpatterned films and the polarization dependence of the plasmon intensity measured in skew ribbon arrays. Such TR- and TA-tunable topological transitions can be well explained based on the effective sheet optical conductivity by adding up those of the two films. Our study demonstrates TR as another degree of freedom for the manipulation of plasmonic topology in nanophotonics, exhibiting promising applications in bio-sensing, heat transfer and the enhancement of spontaneous emission.

**Keywords:** plasmonics, hyperbolic polaritons, two-dimensional materials, photonic topological transition

**TOC Figure:**

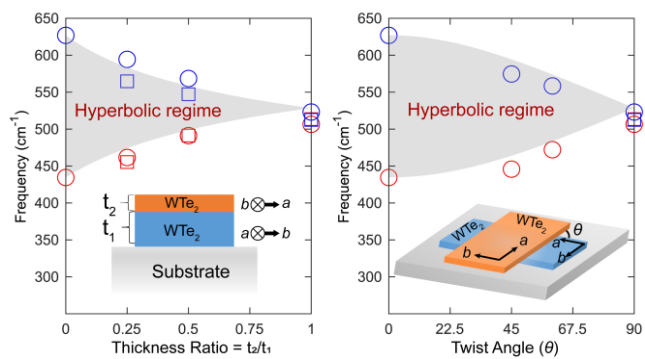

Plasmon polaritons, the coupled modes of free carrier collective oscillations and electromagnetic waves, can induce subwavelength light confinement and significantly boost light-matter interactions for numerous applications[1, 2]. In two-dimensional (2D) films with thickness much smaller than the plasmon wavelength, the topology of the 2D plasmon is determined by the signs of the sheet optical conductivities along two in-plane principal axes, giving rise to elliptic (hyperbolic) iso-frequency contours (IFCs) in the same (opposite) sign configuration[3-6]. Especially in the hyperbolic regime, plasmons have open IFCs, extremely large wavevectors, directional propagation and high optical local density of states, exhibiting great potential for nano-confinement of light[7, 8], polarization engineering[9, 10], enhancement of light-matter interactions[11, 12] and near-field radiative heat transfer[13-16].

The change between such two different plasmonic iso-frequency topologies (i.e., between the elliptic and hyperbolic regimes) is known as photonic topological transition[17], in a close analog to Lifshitz transition in condensed matter physics where the Fermi surface undergoes an evolution from open to closed contours[18]. Tailoring the plasmonic topology between the elliptic and the hyperbolic regimes is crucial for the steering of plasmon propagation[6] and light-matter interactions[19]. The strong light confinement in the ultrathin 2D plasmonic materials with evanescent tails in the immediate environment can facilitate various engineering tools to efficiently tune the plasmon dispersion, such as by modifying the dielectric environment. This can also potentially induce plasmonic topological transitions, which, as an analogous example, has been explored in phonon polaritons of $\alpha$-MoO$_3$, a different but alike quasiparticle involving optical phonons by choosing polar substrates[20, 21], forming heterostructures with graphene[22-26] and stacking twisted bilayer structures[27-33]. In addition to the twist angle (TA) in the twisted bilayer, an unexplored tuning parameter is the thickness ratio (TR) of two layers. When two identical hyperbolic films are stacked with TA = 90° and TR = 1, polaritons in the resulting bilayers behave in the same way as that in an isotropic surface, which has been demonstrated in twisted $\alpha$-MoO$_3$ bilayers[29, 30]. On the other hand, a single film, regarded as TA = 90° and TR = 0, retains the original hyperbolic regime. In the intermediate values of TR, however, whether the hyperbolic regime exists and how the plasmon topology evolves are still elusive. Moreover, it is interesting to check whether there is a unified theory that can account for the tuning effect of both the film thickness ratio and the twist angle.

$T_d$-phase WTe$_2$ is a van der Waals Weyl semimetal, which has been demonstrated to host in-plane hyperbolic plasmon polaritons naturally in the far infrared range at cryogenic temperatures. A plasmonic topological transition occurs when the operation frequency crosses the lower hyperbolic boundary (about 429 cm$^{-1}$ for samples at 10 K)[34], where the exotic propagation features switching from the in-plane ellipticity to hyperbolicity could occur with dramatic modifications of light-matter interaction strength[35]. Here, we stack two WTe$_2$ films with different TR and TA values to showcase the synergistic effect of TR and TA, the two basic geometric parameters, as the new compound degree of freedom for on-demand manipulations of plasmonic topology. Importantly, our methods and findings can be extended to other hyperbolic polaritonic systems.

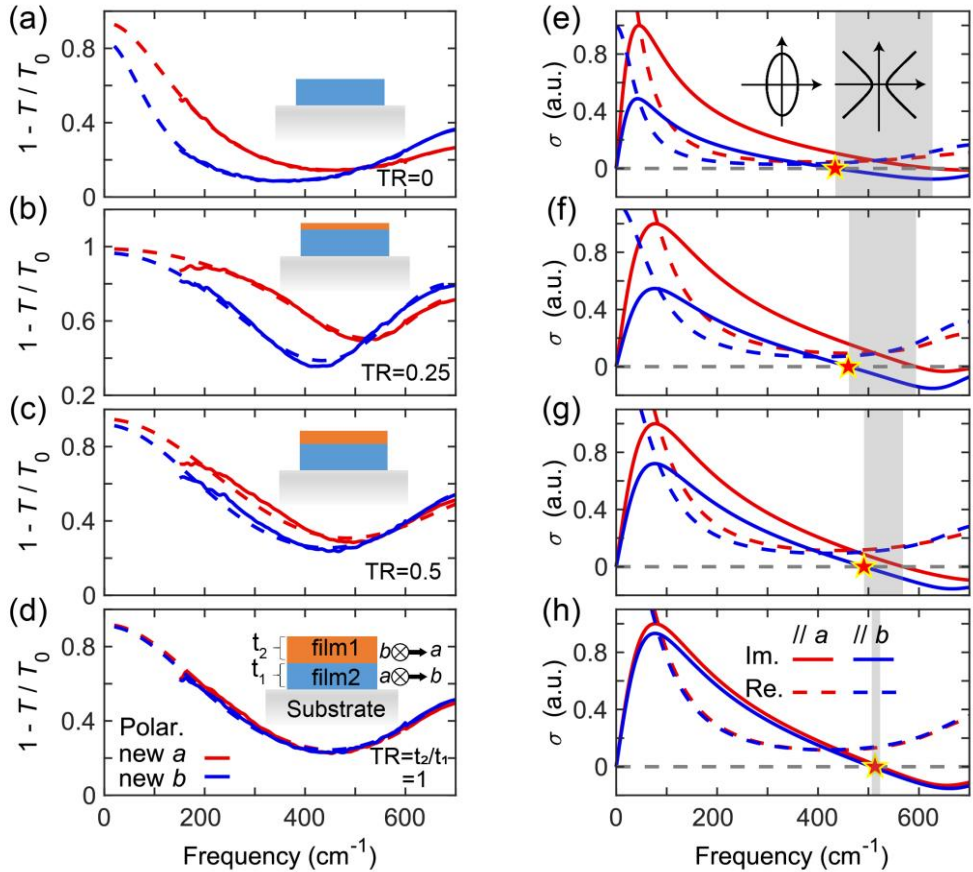

**Figure 1. Extinction spectra of unpatterned WTe₂ bilayers.** (a) to (d), Measured anisotropic extinction spectra with different TR values and a fixed TA of 90° (solid curves). The dashed curves are the corresponding fittings (see Supporting Information for fitting details). Insets show the schematics of the 90°-twisted WTe₂ bilayers. (e) to (h) Fitted real (dashed lines) and imaginary (solid lines) parts of conductivities along two new axes for different TR values, with the zero crossing points along *b* axis marked by red stars. Hyperbolic regimes are indicated by gray shaded areas. IFCs with different topologies are illustrated by insets in (e).

We first examine the effective sheet optical conductivity of unpatterned twisted bilayers with different TRs but a fixed TA of 90° from the infrared extinction spectra. The bilayer structure fabrication process is shown in the Supporting Information. Figures 1a to 1d display the measured extinction spectra of the 90°-twisted WTe₂ bilayers with different TR values (see Supporting Information for measurement details). Here, the extinction spectrum is defined by $1 - T/T_0$, where $T$ and $T_0$ are the transmission of light through the film and the bare substrate respectively, and TR (restricted in 0 to 1) represents the thickness ratio between the top (thinner) and bottom (thicker) films as shown in insets of Figures 1a to 1d. Note that there is no spacer layer between the two WTe₂ films. We define *a* (*b*) axis of the bottom films as the new *a* (new *b*) axis of the bilayers with TA of 90°, since the bottom films are always thicker in our experiments. All the measurements were performed at 10 K. As shown in Figures 1a to 1d, the anisotropy dwindles at higher TR, as manifested by the reduced difference between the extinction spectra polarized along two new principal axes. In Figure 1d, the spectrum shows nearly no polarization dependence when TR = 1. In view of the ultra-thin thickness of the stacked bilayer (~100 nm) compared to the wavelength (>15 μm), we treat it as a homogeneous film with its own effective sheet optical conductivity, which can be extracted from the extinction spectra. Figures 1e to 1h show the corresponding extracted imaginary parts of the anisotropic optical conductivity for the bilayer films in Figures 1a to 1d (see Supporting Information for fitting details). Based on the signs of the conductivities along two axes, we can identify the plasmon topology of the bilayer (see Note III of Supporting Information for more details). The hyperbolic regime, denoted as the shaded area, shrinks as TR increases, and the associated topological transition to elliptical regime (indicated by red stars) blueshifts in frequency. Extinction spectra of twisted bilayers with different TA (0° to 90°) and fixed TR=1 (the same thickness for the top and bottom films) are shown in Figure S3 in Supporting Information. TA-induced modulation of the hyperbolic regime can be observed, which was also realized previously for phonon polaritons in twisted α-MoO₃ bilayers[30-33].

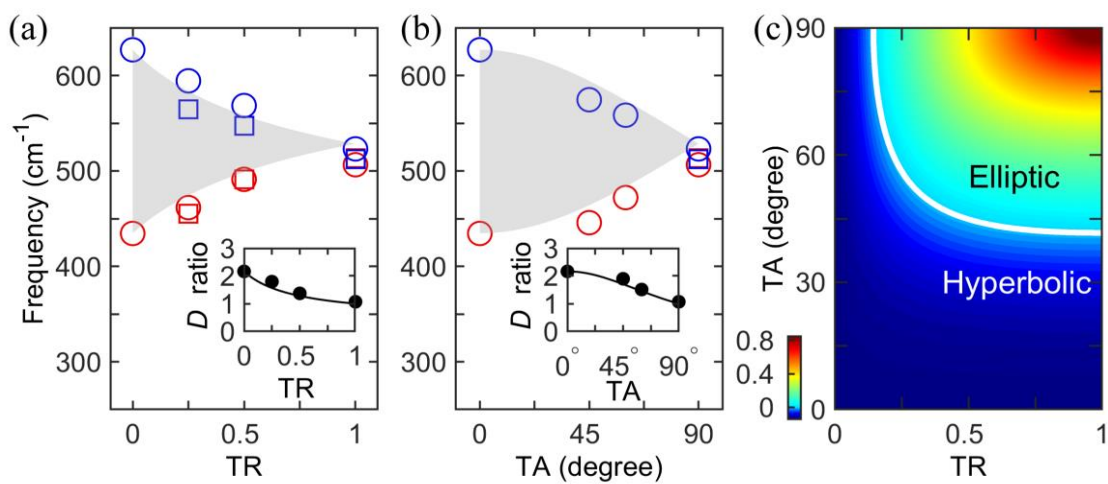

**Figure 2. Evolution of plasmonic topology in unpatterned WTe₂ bilayers.** (a) and (b) Evolution of hyperbolic regimes as a function of TR and TA values, with TA in (a) and TR in (b) fixed at 90° and 1, respectively. Blue and red spots represent the upper and lower hyperbolic boundaries determined from the fitted conductivities of extinction spectra of bilayers. The upper and lower boundaries fitted from one film are plotted as the same symbols, such as circles or squares. Hyperbolic regimes calculated based on Eq. (1) are indicated by shaded areas. Insets in (a) and (b) exhibit evolution of the Drude weight ratio between the new *a* and new *b* axes. (c) Diagram of the calculated $\sigma''_{\text{eff } bb}/\sigma''_{\text{eff } aa}$ at 460 cm⁻¹ based on Eq. (1). The white curve represents the boundary between elliptic and hyperbolic regimes with $\sigma''_{\text{eff } bb}/\sigma''_{\text{eff } aa} = 0$.

The hyperbolic boundaries determined by the extracted conductivities in unpatterned bilayers with fixed TA (TR) are summarized in Figure 2a (Figure 2b) as a function of TR (TA). The lower and upper boundaries are plotted in red and blue spots respectively. Note that cases for (TR = 0, TA = 90°) in Figure 2a and (TR = 1, TA = 0°) in Figure 2b share the same hyperbolic regime, since they both regress to a bare single WTe$_2$ film. In thin bilayers with total thickness much smaller than the free-space light wavelength and the plasmon wavelength, the optical properties of the new system can be captured by the effective sheet optical conductivity tensor $\sigma_{eff}(\omega)$ by summing the tensors of the bottom and top layers weighted by the film thickness (see Note I in the Supporting Information for details)[27-29]. Thus, we have:

$$\sigma_{eff}(\omega) = t_1 \cdot \sigma^{3D}(\omega) + t_2 \cdot \widetilde{\sigma^{3D}}(\omega) \qquad (1)$$

where $\sigma^{3D} = \begin{bmatrix} \sigma_{aa}^{3D} & 0 \\ 0 & \sigma_{bb}^{3D} \end{bmatrix}$ is the bulk conductivity tensor, and a rotated tensor $\widetilde{\sigma_{3D}} =$

$\begin{bmatrix} \sigma_{aa}^{3D}\cos^2\theta + \sigma_{bb}^{3D}\sin^2\theta & \sin\theta\cos\theta\,(\sigma_{bb}^{3D} - \sigma_{aa}^{3D}) \\ \sin\theta\cos\theta\,(\sigma_{bb}^{3D} - \sigma_{aa}^{3D}) & \sigma_{aa}^{3D}\sin^2\theta + \sigma_{bb}^{3D}\cos^2\theta \end{bmatrix}$ is used for the top film with a twist angle $\theta$; $t_1$ and $t_2$ are the thickness of the bottom and top films respectively, with TR = $t_2/t_1$. The shaded areas in Figures 2a and 2b show the calculated hyperbolic regimes based on Eq. (1), with $\sigma^{3D}$ obtained from the sheet optical conductivity of a single film in Figure 1e normalized by the corresponding film thickness. The hyperbolic boundaries of $\sigma_{eff}(\omega)$ are determined by zeros of the imaginary parts along two new principal axes. As shown in Figures 2a and 2b, the shaded areas are consistent with the measured data points. The lower and upper hyperbolic boundaries blue- and red- shift, respectively, upon the increasing of TR or TA. They finally merge for the case of (TR = 1, TA = 90°) at the frequency where $\sigma_{aa}'' + \sigma_{bb}'' = 0$ in single WTe$_2$ films. Moreover, from the extinction spectra in the bilayers, we can get the Drude weights. The ratio of the fitted Drude weights along the new $a$ and new $b$ axes are displayed in insets of Figures 2a and 2b, showing good consistency with calculated results (black lines) based on Eq. (1). The ratio decreases and approaches to 1 at higher TR or TA values, suggesting the reduced anisotropy of the bilayer.

With Eq. (1) and the experimentally obtained $\sigma^{3D}$, the plasmon topology at a specific frequency for bilayers with various TR and TA could be depicted. For instance, the evolution of the ratio $\sigma_{eff\,bb}''/\sigma_{eff\,aa}''$ with TR and TA at 460 cm$^{-1}$ are displayed in Figure 2c, where $\sigma''_{\text{eff}\,aa}$ and $\sigma''_{\text{eff}\,bb}$ are the imaginary parts of the diagonal elements of $\sigma_{\text{eff}}$. This serves as a topology diagram in the TR and TA parameter space. At small TA and TR values, $\sigma''_{\text{eff}\,bb}/\sigma''_{\text{eff}\,aa}$ is negative, indicating the hyperbolic topology. As TA and TR increase and cross the white curve, at which $\sigma''_{\text{eff}\,bb}/\sigma''_{\text{eff}\,aa} = 0$, a topological transition occurs, demonstrating the capability of topology manipulation both by TA and TR. Such transition occurs at higher TA and TR values for higher plasmon frequencies, consistent with the results in Figures 2a and 2b (see Figure S5, S8 to S10 in the Supporting Information about the calculation of the plasmon dispersion and related topological transitions.).

The above extinction spectra of unpatterned bilayers give us valuable information on the effective optical conductivity and infer the manipulation of plasmon topology. Now let's directly interrogate the topology by exciting plasmons in patterned structures. To fulfill this, we investigated localized plasmons of skew ribbon arrays. In the experiment, the ribbon width is much larger than the thickness of the bilayer. Schematics of the skew ribbons are shown in Figures 3a and 3c, where ribbon directions are away from the new principal axes. In contrast to the isotropic surface where the plasmon intensity is strongest with light polarized perpendicular to the ribbons, the optimal polarization for anisotropic surface deviates from the perpendicular direction of skew ribbon arrays[36]. As demonstrated by theory[37] and simulations (see Note II and Figure S11 in Supporting Information for details), the optimal polarization angle $\Phi_{max}$ is determined by the ratio of the imaginary parts of the sheet optical conductivities along two principal axes with the relation $\tan(\Phi_{max})/\tan(\theta) = \sigma''_{\text{eff}\,bb}/\sigma''_{\text{eff}\,aa}$, where $\theta$ is the skew angle for ribbons, as shown in Figure 3c. Note that all the skew ribbon arrays in this work have negative skew angles ($\theta < 0$). Thus, the plasmon topology can be determined by the sign of $\Phi_{max}$, i.e., elliptic when $\Phi_{max} < 0$, and hyperbolic when $\Phi_{max} > 0$, as illustrated in Figures 3a and 3c.

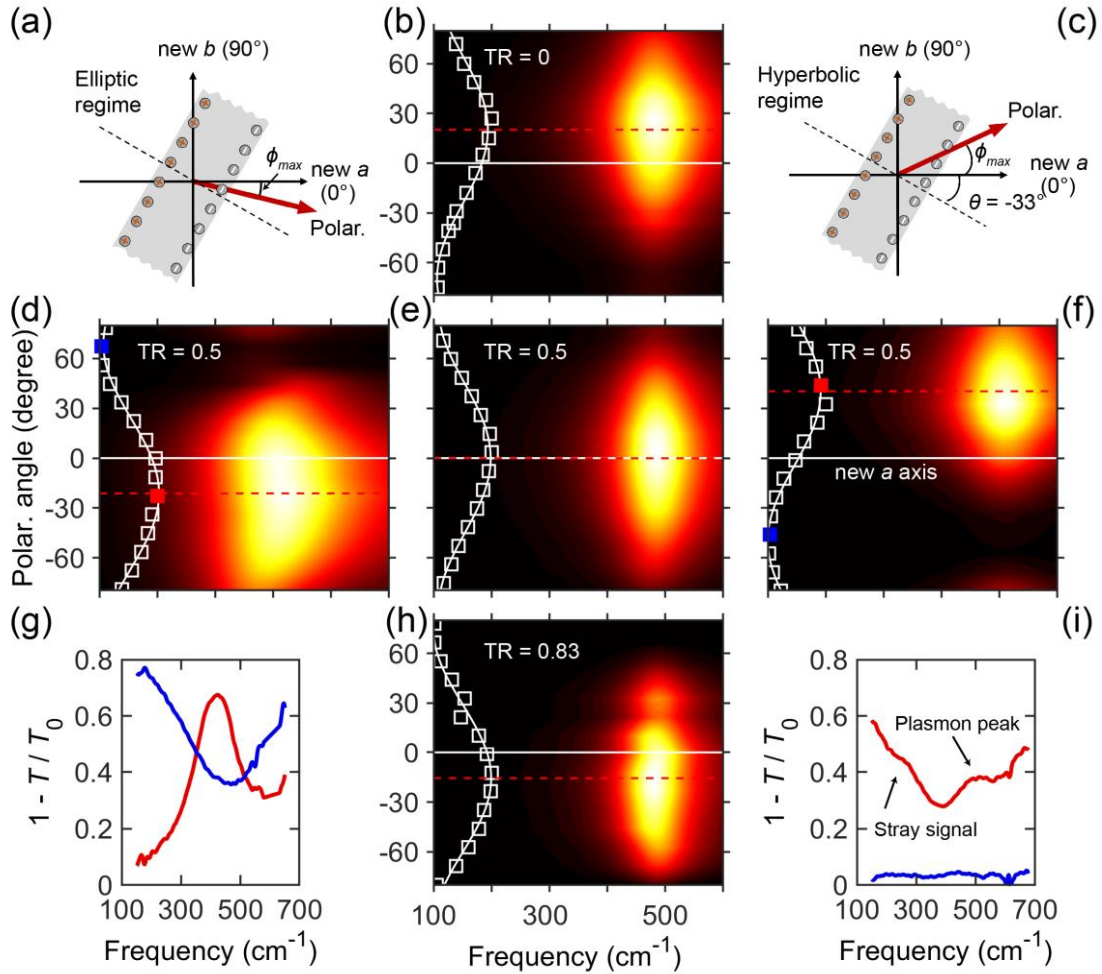

**Figure 3. Polarization dependence of the plasmon in skew ribbon arrays.** (a) and (c) Schematics of skew ribbon arrays in hyperbolic and elliptic regimes. (b) (d) (e) (f) and (h) Polarization dependence of plasmon intensity measured in skew ribbon arrays of bilayers of TA = 90° with different TR values. The maximum values are normalized to 1. See table S1 in Supporting Information for details of film thickness and ribbon width. All skew angles are -33°. White squares in insets are the fitted plasmon intensity evolution as a function of the polarization. Red dashed lines denote the polarization angles for plasmon intensity maximum. Red and blue squares in (d) and (f) denote the polarizations for maximal and minimal plasmon intensity. (g) and (i) Measured extinction spectra for polarization with maximal and minimal plasmon intensity for skew ribbon arrays in (d) and (f). The little bump around 300 cm$^{-1}$ in (i) is the stray signal from a nearby ribbon array.

With this convenient indicator of the plasmon topology, we first reveal the evolution of the topology as a function of the resonance frequency in bilayers with fixed TR or TA values. Figures 3d to 3f show the polarization dependence of the extracted plasmon resonance spectra (an example of the series of raw spectra are displayed in Figure S6 of Supporting Information) in skew ribbon arrays with $\theta = -33°$, patterned from 90°-twisted WTe$_2$ bilayers with TR of 0.5. Here, the Drude response and interband transitions have been removed to highlight the variations in the plasmon resonances as a function of the polarization angle. Fitting of the extinction spectrum of a typical skew ribbon array is shown in Figure S12 of Supporting Information (see Supporting Information for fitting details). Those three ribbon arrays show different plasmon resonance frequencies, mainly determined by the predesigned ribbon width. From Figures 3d to 3f, with increasing plasmon frequency, the optimal polarization (marked by red dashed lines) changes from negative to positive values, suggesting an elliptic to hyperbolic topological transition. The raw extinction spectra at the polarization near the maximal and minimal plasmon intensities in Figures 3d and 3f (red and blue squares) are showed as red and blue lines in Figures 3g and 3i, respectively. In Figure 3g, the red curve has the maximal intensity of plasmon and in the meantime the Drude response is nearly negligible, which is similar to that in isotropic surfaces such as graphene[38]. While in Figure 3i, the Drude response is prominent at the polarization for the maximal plasmon intensity (the red curve) but nearly absent for the minimal plasmon intensity (the blue curve), which is a result of the extreme anisotropy in the hyperbolic regime. The plasmon resonance of the bilayer in Figure 3d has a frequency of 410 cm$^{-1}$, near the lower hyperbolic boundary (429 cm$^{-1}$) for single WTe$_2$ films where $\Phi_{max}$ is 0°. However, the fitted $\Phi_{max}$ for Figure 3d is -21.3°, with a significant deviation from 0°, indicating the decreased optical anisotropy after 90°-twisted stacking at this frequency. Plasmons in Figure 3e with resonance frequency of 486 cm$^{-1}$ have maximal intensity at polarization along the new $a$ axis ($\Phi_{max} = 0.002°$), implying that the lower hyperbolic boundary has blueshifted at higher TR values, changing from 429 cm$^{-1}$ for TR = 0 and TA = 90° to about 486 cm$^{-1}$ for TR = 0.5 and TA = 90°, consistent with the result in Figure 2a.

Next, we check the effect of TR on tuning the plasmon topology in more detail. Figures 3b and 3h shows the results of skew ribbon arrays with $\theta = -33°$ for different TR values with TA of 90°. They have nearly the same plasmon resonance frequency ($\pm 2$ cm$^{-1}$) as that in Figure 3e. In Figure 3b (TR = 0), plasmons have the maximal intensity at $\Phi_{max} = 15.6°$, while when TR increases to 0.83, $\Phi_{max}$ is -15.5°, as shown in Figure 3h. Thus, as TR increases from 0 to 0.83 in Figures 3b, 3e and 3h, the plasmon topology changes from the hyperbolic to the elliptic, demonstrating TR-induced tuning of the plasmon topological transition in 90°-twisted WTe₂ bilayers.

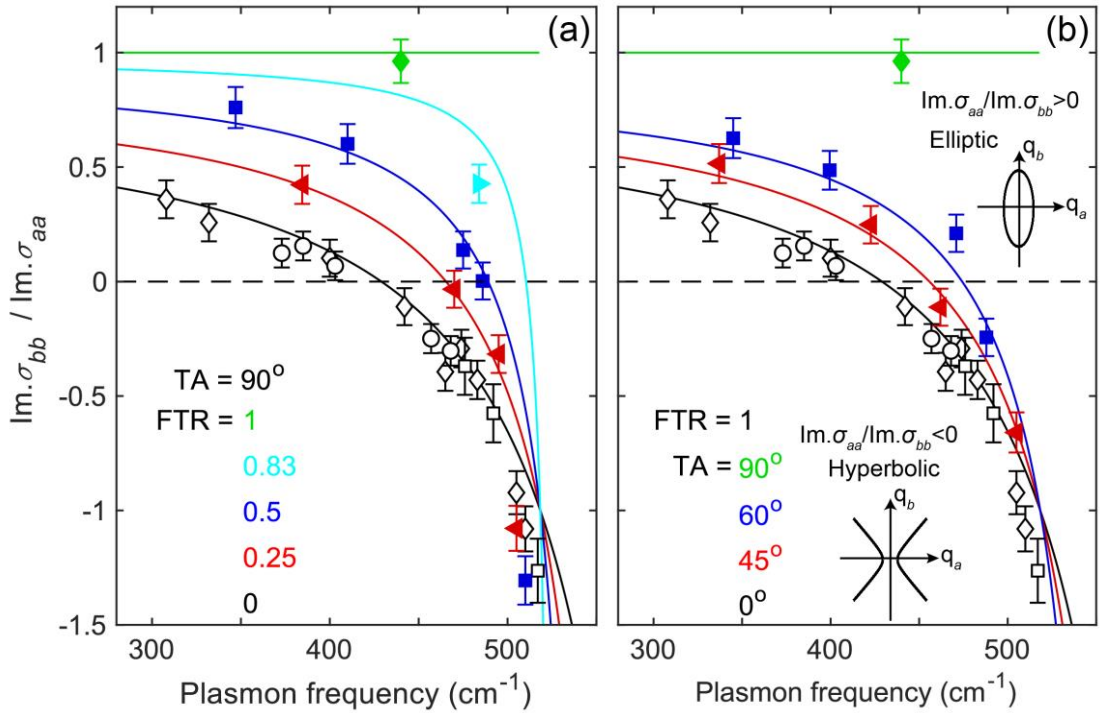

**Figure 4. Imaginary parts of effective conductivities $\sigma''_{\mathrm{eff}\,bb}/\sigma''_{\mathrm{eff}\,aa}$ in different geometric configuration:** (a) and (b) Evolution of ratio of $\sigma''_{\mathrm{eff}\,bb}/\sigma''_{\mathrm{eff}\,aa}$ as a function of plasmon frequency in bilayers with different TR (a) and TA (b) values. Data points are calculated by $\Phi_{max}$ in skew ribbon arrays. Black hollow points in (a) for TR = 0 and (b) for TA = 0° are measured in ribbons with skew angle of -23° (square points), -33° (diamond points), -40° (circle points). Other data points are measured in ribbons with skew angle of -33°. Points in (a) and (b) have TR and TA deviation from nominal values in the range of 0.04 and 1° respectively. The solid lines are calculated

by summing the top and bottom sheet conductivities based on conductivities determined from the plasmon dispersion in Ref. [34].

As mentioned above, the sign of the ratio $\sigma''_{\text{eff} \, bb}/\sigma''_{\text{eff} \, aa}$ determines the topology of plasmons. More quantitatively, this ratio can be obtained experimentally from $\theta$ and $\Phi_{max}$ based on $\tan(\Phi_{max})/\tan(\theta) = \sigma''_{\text{eff} \, bb}/\sigma''_{\text{eff} \, aa}$, the latter angle of which is determined by fitting the polarization dependence of the plasmon intensity. Data points in Figures 4a and 4b show thus obtained $\sigma''_{\text{eff} \, bb}/\sigma''_{\text{eff} \, aa}$ for bilayers with different TR (fixed TA = 90°) and TA (fixed TR = 1), respectively. For single WTe$_2$ films, which correspond to cases of (TR = 0, TA = 90°) and (TR = 1, TA = 0°), data points from different skew angles ($\theta = -23°, -33°, -40°$ are plotted together in Figures 4a and 4b. Other data points for bilayers with higher TR and TA in Figures 4a and 4b are all measured in skew ribbons with $\theta$ of $-33°$. The solid curves in Figures 4a and 4b are calculated from the effective conductivity tensors by summing sheet conductivities of the top and bottom films (Eq. (1)) using values determined by the plasmon dispersion in Ref. [34] (See Supporting Information for details). As shown in Figures 4a and 4b, data points can be well described by the solid lines, indicating that the effective conductivity tensors work well in plasmonic bilayer structures. In Figure 4a (4b), as TR (TA) increases, the zero point of the conductivity ratio (crossing with the black dashed line) moves to higher frequencies, demonstrating that the lower hyperbolic boundary can be successfully tuned by changing TR (TA). Note that all solid curves in Figures 4a and 4b, except the case for (TR = 1, TA = 90°), cross atf the same point at around 518 cm$^{-1}$. This is due to the fact that $\sigma''_{aa} + \sigma''_{bb} = 0$ at this frequency in single WTe$_2$ films, which renders $\sigma''_{\text{eff} \, bb}/\sigma''_{\text{eff} \, aa}$ equal to -1 regardless of TR and TA values. Such convergence can be also inferred from data points in Figure 4a. Fitting curves using sheet conductivities extracted from the extinction spectra of the bare film in Figure 1e give consistent results with data points in Figures 4a and 4b as well (see Figure S7 of Supporting Information for details). Therefore, a simple addition of film optical conductivities for the stacked bilayer works well in the interpretation of our experimental findings.

Our work provides a new avenue to engineer the effective conductivity tensor by stacking different films with zero or thin spacing layer and hence to tune the plasmonic functionalities and light-matter interactions. Our experiment demonstrates that TR and TA have comparable effects on tuning plasmon topology in bilayer structures. In fact, tuning thickness ratio is equivalent to tuning the effective sheet conductivity tensor, which gives us a universal method to tailor the topology of homo- or hetero-structures. Note that, however, when the plasmon wavelength is comparable to the film thickness ($t$) or the distance between the two films ($d$), the coupling between the two films will be weakened, and a simple sum of conductivities in Eq. (1) to describe the plasmon response tends to fail. For plasmon modes with extremely large wave vectors and the top and bottom layers are separated slightly by a spacer, the coupling goes to zero, leaving the two plasmonic surfaces uncoupled, as suggested in Ref. [29]. Future efforts can be devoted to the tuning of the film distance with a spacer.

In summary, we demonstrated the TR- and TA-tuning of plasmon topology in twisted WTe$_2$ bilayer system from 16 μm to 23 μm in wavelength by measuring the film extinction spectra and the polarization dependence of the plasmon in skew ribbon arrays. The plasmon topology is demonstrated to be directly governed by the effective conductivity tensor obtained by a sum of the respectively rotated conductivity tensors of the top and bottom layers. The topology manipulation of polaritons in such long wavelength will facilitate many promising applications in far-IR and THz range, such as polarization engineering, enhancement of spontaneous emission, thermal emission management and biosensing.

ASSOCIATED CONTENT

**Supporting Information:**

Detailed description of the effective sheet optical conductivity method, topological transitions of 2D plasmons, device fabrication and optical image, extinction spectra of WTe2 bilayers with different TWs, polarization dependence of raw extinction spectra of plasmons, fitting results and uncertainties of the extinction spectra in bare films and ribbon arrays, plasmon dispersion as function of carrier density, loss function calculation of the effect of TR and TA on plasmon topology, simulations for the polarization dependence of plasmon intensity in skew ribbon arrays, film thickness and ribbon width.

AUTHOR INFORMATION

**Corresponding Author**

*Hugen Yan

State Key Laboratory of Surface Physics, Key Laboratory of Micro and Nano-Photonic Structures (Ministry of Education), and Department of Physics, Fudan University, Shanghai 200433, China.

Email: hgyan@fudan.edu.cn

*Yugui Yao

Centre for Quantum Physics, Key Laboratory of Advanced Optoelectronic Quantum Architecture and Measurement (MOE), School of Physics, Beijing Institute of Technology, Beijing, 100081, China.

Email: ygyao@bit.edu.cn

* Cheng-Wei Qiu

Department of Electrical and Computer Engineering, National University of Singapore, Singapore 117583, Singapore

Email: eleqc@nus.edu.sg

**Author Contributions**

H.Y., Y.Y. and Q.C. initiated the project and conceived the experiments and theory.

C.W. and X. Y. prepared the samples, performed the measurements and data analysis with assistance from H.Y., M.J., H.S., S.C., W.F., L.Y., Z.J. and M.L.. C.W. carried out the fittings with the help from H.G. and X.Q.. H.Y., C.W. and X.Y. co-wrote the manuscript. H.Y. supervised the whole project. All authors commented on the manuscript.

ACKNOWLEDGEMENT

H.Y. is grateful to the financial support from the National Key Research and Development Program of China (Grant Nos. 2021YFA1400100, 2022YFA1404700), the National Natural Science Foundation of China (Grant No. 12074085), the Natural Science Foundation of Shanghai (Grant No.23XD1400200).

Y.Y. is supported by the National Key Research and Development Program of China (Grant No. 2020YFA0308800), the National Natural Science Foundation of China (Grants Nos. 12061131002，12234003)，the Strategic Priority Research Program of Chinese Academy of Sciences (Grant No. XDB30000000).

C.W. is grateful to the financial support from National Key Research and Development Program of China (Grant No. 2022YFA1403400) and the National Natural Science Foundation of China (Grant Nos. 12274030, 11704075).

C.-W.Q. acknowledges the support from the National Research Foundation, Prime Minister's Office, Singapore, under its Competitive Research Programme (CRP award NRF CRP26-2021-0004)

Q.X is grateful to the financial support from the China Postdoctoral Science Foundation (Grant No. KLH1512129).

S.H. is grateful to the financial support from the China Postdoctoral Science Foundation (Grant No. 2020TQ0078).

Y.H. is grateful to the National Key Research and Development Program of China (Grant No. 2019YFA0308000), the National Natural Science Foundation of China (Grant No. 62022089) and Chongqing Outstanding Youth Fund (Grant No. 2021ZX0400005).

Part of the experimental work was carried out in Nanofabrication Lab of Fudan

University and Beijing Institute of Technology.

ABBREVIATIONS:

TA, twist angle; TR, thickness ratio; IFCs, iso-frequency contours; AFM, atomic force microscope; FTIR, Fourier transform infrared.